\documentstyle[12pt]{article}

\newfont{\twelvemsb}{msbm10 scaled\magstep1}
\newfont{\eightmsb}{msbm8}
\newfam\msbfam
\textfont\msbfam=\twelvemsb
\scriptfont\msbfam=\eightmsb
\catcode`\@=11
\def\Bbb{\ifmmode\let\next\Bbb@\else
  \def\next{\errmessage{Use \string\Bbb\space only in math mode}}\fi\next}
\def\Bbb@#1{{\fam\msbfam{{#1}}}}

\newcommand{\be}{\begin{equation}}
\newcommand{\ee}{\end{equation}}
\newcommand{\ba}{\begin{eqnarray}}
\newcommand{\ea}{\end{eqnarray}}
\newcommand{\spz}{\hspace{0.5cm}}
\newcommand{\virg}{\spz,\spz}

\newcommand{\resection}[1]{\setcounter{equation}{0}\section{#1}}

\def\d{\delta}

\def\t{\theta}

\newcommand{\la}{\lambda}
\newcommand{\p}{\partial}
\newcommand{\dx}{\partial_x}
\newcommand{\dt}{\partial_t}
\newcommand{\dla}{\partial_{\lambda}}
\newcommand{\half}{{\textstyle\frac{1}{2}}}

\newcommand{\nn}{\nonumber}
\newcommand{\lt}{\left(}
\newcommand{\rt}{\right)}

\newcommand{\buno}{\mbox{\bf 1}}

\newcommand{\cL}{{\cal L}}

\newcommand{\NP}[1]{Nucl.\ Phys.\ {\bf #1}}
\newcommand{\PL}[1]{Phys.\ Lett.\ {\bf #1}}

\newcommand{\CMP}[1]{Comm.\ Math.\ Phys.\ {\bf #1}}
\newcommand{\CPAM}[1]{Comm.\ Pure\ Appl.\ Math.\ {\bf #1}}
\newcommand{\PR}[1]{Phys.\ Rev.\ {\bf #1}}
\newcommand{\PRL}[1]{Phys.\ Rev.\ Lett.\ {\bf #1}}
\newcommand{\MPL}[1]{Mod.\ Phys.\ Lett.\ {\bf #1}}

\newcommand{\LMP}[1]{Lett.\ Math.\ Phys.\ {\bf #1}}

\begin{document}
\sloppy
\renewcommand{\thefootnote}{\fnsymbol{footnote}}

\newpage
\setcounter{page}{1}

\vspace{0.7cm}
\begin{flushright}
46/2000/EP\\
April 2000
\end{flushright}
\vspace*{1cm}
\begin{center}
{\bf  Hidden Virasoro Symmetry of (Soliton Solutions of) the Sine Gordon
Theory} 
\\ \vspace{1.8cm}
{\large D.\ Fioravanti $^a$\footnote{E-mail:
dfiora@sissa.it} and M.\ Stanishkov $^b$\footnote{E-mail:
stanishkov@bo.infn.it ; on leave of absence from I.N.R.N.E. 
Sofia, Bulgaria}}\\  \vspace{.5cm} $^a${\em S.I.S.S.A.-I.S.A.S. and I.N.F.N.
Sez. di Trieste \\      Via Beirut 2-4 34013 Trieste,
Italy} \\ $^b${\em Dipartimento di Fisica, Universita' di Bologna\\  Via
Irnerio 46, 40126 Bologna, Italy} \\  \end{center}
\vspace{1cm}

\renewcommand{\thefootnote}{\arabic{footnote}}
\setcounter{footnote}{0}

\begin{abstract}
{\noindent We present} a construction of a Virasoro symmetry of the
sine-Gordon ( SG ) theory. It is a dynamical one and has nothing to do with
the space-time Virasoro symmetry of 2D CFT. Although it is clear how it can be
realized directly in the SG field theory, we are rather concerned here with
the corresponding N-soliton solutions. We present explicit expressions for
their infinitesimal transformations and show that they are local in this
case. Some preliminary stages about the quantization of the classical results
presented in this paper are also given.

\end{abstract}

\vspace{1cm}

\newpage

\resection{Introduction}

The 2D sine-Gordon model, defined by the action:
\be
S={\pi\over\gamma}\int {\cL}d^2x \virg {\cL}=(\p_\nu\phi)^2-m^2(cos(2\phi)-1),
\ee
where $\gamma$ is the coupling constant and $m$ is related to the mass scale,
is one of the simplest Integrable Quantum Field
Theories. It possesses an infinite number of conserved charges $I_{2n+1}$,
$n\in {\Bbb Z}$, in involution. At present, much is known about the
corresponding scattering theory. It contains solitons, antisolitons and a
number of bound states called ``breathers''. The mass spectrum and the
S-matrix have been known for about 20 years \cite{COL}. Despite this on-shell
information, the off-shell Quantum Field Theory is much less developed. In
particular, the computation of the corresponding correlation functions is
still an important open problem. Actually, some progress towards this
direction has been made recently. For instance, the exact Form-Factors ( FF's)
of the exponential fields  $<0|\exp\phi(0)|\beta_1,...,\beta_n>$ were computed
\cite{SL}. This allows one to make predictions about the long-distance
behaviour of the corresponding correlation functions. On the other hand, some
efforts have been made to estimate the short distance behaviour of the theory
in the context of the so-called Conformal Perturbation Theory (CPT)
\cite{ALZ}. By combining the previuos techniques (FF's and CPT), it has been
possible to extimate several interesting physical quantities (\cite{FMS} and
references therein). In addition, the exact expression for the Vacuum Expectation Values (VEV's) of
the exponential fields ( and some descendents ) were proposed in \cite{LFZZ}.
The  VEV's provide a highly non-trivial non-perturbative information about the
short distance expansion of the two-point correlation functions. The
FF's and CPT  approaches permit to make predictions about the approximate
behaviour  of the correlation functions of the sine-Gordon theory in the
infrared and ultraviolet regions correspondingly. What remains still unclear
however is the explicit form of the correlation functions, in particular their
intermediate behaviour and analytic properties, a question of primary
importance from the field-theoretical point of view. Some exact results exist
only at the so-called free-fermion point $\gamma={\pi\over 2}$ \cite{MCCOY}.

There exists another approach to the sine-Gordon theory. It  consists in
searchig for additional infinite-dimentional symmetries and is inspired by the
success the latter had in the 2D CFT. In fact, it has been shown in \cite{SBL}
that the sine-Gordon theory possesses an infinite dimensional symmetry
provided by the $\widehat sl(2)_q$ algebra. However, this symmetry connects the
correlation functions of the fields in the same multiplet
without giving a sufficient information about the functions
themselves. 

It is known to some extent that there should be another kind of symmetry
present in the sine-Gordon theory. Actually, it is known that it can be
obtained as a particular scaling limit of the so-called XYZ -- spin chain
\cite{LUT}.  The
latter is known to possess an infinite symmetry obeying the so-called Deformed
Virasoro Algebra (DVA)\cite{LUKY}.  It is natural to suppose that 
in the scaling limit, represented by SG, there should be present some infinite
dimensional symmtery, a particular limit of DVA. At present, a lot is known
about the mathematical structure of DVA, in particular the highest weight
representations and the screening charges have been constructed
\cite{LUKJAP}. What remains unclear is how the corresponding symmetry is
realized in the sine-Gordon field theory, for example what is the action of
the corresponding generators on the exponential fields, what kind of
restrictions it imposes on their correlation functions etc.

In this paper we present a construction of a Virasoro symmetry directly in the
sine-Gordon theory. Although we are of course interested in the quantum
theory, we restrict ourself to the classical picture in this paper. Also,
though it will be clear how to implement it in the general field theory, we
are mainly concerned here with the construction of this symmetry in the case of
the N-soliton solutions.  One of the reasons for this is that the symmetry in
this case is much simpler realized ( in particular it becomes {\it local}
contrary to the field theory realization ). We were also inspired by the work
of Babelon,Bernard and Smirnov \cite{BB}. It was shown there that certain
form-factors can be directly reconstructed by a suitable quantization of the
N-soliton solutions. It was also shown in \cite{BBS} that certain null-vector
constraints arise in sine-Gordon theory leading to integral equations for the
corresponding form-factors. It is an intriguing question of what is the
symmetry structure lying behind it and in particular its relation to the
Virasoro symmetry we present here.

This paper is organised as follows. In the next Section we recall the
construction of the so-called additional non-isospectral Virasoro symmetry of
KdV theory. It is done in the context of the so-called algebraic approach and
is a generalization of the well known dressing symmetries of integrable models
\cite{FS}. In Section 3 we explain how one can restrict this symmetry to the
case of N-soliton solutions of KdV. It happens that it becomes {\it local} in
this case, contrary to the field theory realization where it is
{\it quasi-local}. In Section 4 we extend the Virasoro symmetry to
the sine-Gordon soliton solutions. This is acheeved by introducing an
additional {\it time} dynamics in the KdV theory. In such a way we obtain
``negative'' Virasoro flows which complete the ``positive'' ones of KdV to the
whole Virasoro algebra in the sine-Gordon theory. Finally,
in Section 5 we summarise the results obtained in the paper. We give some
hints about the quantization of the classical picture presented here and
discuss some important open problems.

\resection{Virasoro Symmetry of mKdV theory}

\subsection{Dressing Symmetries}

Let us recall the construction of the Virasoro symmetry in the context
of (m)KdV theory\cite{OS, FS1}.  It was shown in \cite{FS}, following the so
called algebraic approach, that it appears as a generalization of the ordinary
dressing transformations of integrable models. Here we briefly recall 
the main results of this article.
Being integrable, the mKdV system admits a zero-curvature representation:
\be 
[ \dt - A_t , \dx - A_x ] = 0 ,
\label{zcurv}
\ee
where the Lax connections $A_x$ , $A_t$ belong to $A_1^{(1)}$ loop algebra:
\ba
A_x &=& - v h + (e_0 + e_1),  \nonumber \\
A_t &=& \la^2(e_0 + e_1 - vh) -
\frac{1}{2}[(v^2-v')e_0 + (v^2+v')e_1] - \frac{1}{2}(\frac{v''}{2}-v^3)h
\label{lax}
\ea
( $v$ is connected to the mKdV field $\phi$: $v=-\phi'$ ) and
\be
e_0=\left(\begin{array}{cc} 0 & \la \\
                                  0 & 0 \end{array}\right)=\lambda E\virg
e_1=\left(\begin{array}{cc} 0 & 0 \\
                                  \la & 0 \end{array}\right)=\lambda F\virg
h  =\left(\begin{array}{cc} 1 & 0 \\
                               0 & -1 \end{array}\right)=H 
\label{gen}
\ee
are the corresponding generators in the fundamental representation.  The usual
KdV variable $u$ is connected to the mKdV field $\phi$ by the Miura
transformation:
\be
u={1\over 2}(\phi')^2+{1\over 2}\phi''
\label{miura}
\ee
(we denote by prime the derivative with respect to the {\it space variable} $x$
of KdV). Of great importance in our construction is the solution
$T(x,\lambda)$ to the so called associated linear problem:
\be
{\cal L}T(x,\lambda)\equiv(\partial_x-A_x(x,\lambda))T(x,\lambda)=0
\label{aslin}
\ee
which is usually referred to (with suitable normalization) as a transfer
matrix. A formal solution to (\ref{aslin}) can be easily found:
\ba 
T_{reg}(x,\la)=e^{H\phi(x)}{\cal P} \exp\lt\la \int_0^xdy
(e^{-2\phi(y)} E+ e^{2\phi(y)} F )\rt = \nn \\
=\left(\begin{array}{cc} A & B \\
                         C & D \end{array}\right).
\label{regsol} 
\ea
It is obvious that this solution defines $T(x,\lambda)$ as an infinite series
in positive powers of $\lambda$ with an infinite radius of convergence (we
shall often refer to (\ref{regsol}) as {\it regular expansion}). For further
reference we present also the expansion of the corresponding matrix elements:
\ba
A=e^{\phi}(1+\sum_1^{\infty}\lambda^{2n}A_{2n})\virg
B=e^{\phi}\sum_0^{\infty}\lambda^{2n+1}B_{2n+1}, \nn \\
 C[\phi]=B[-\phi]\virg D[\phi]=A[-\phi]. 
\label{regmatrix}
\ea
It is also clear from (\ref{regsol}) that $T(x,\la)$ possesses an essential
singularity at infinity where it is governed by the corresponding asymptotic
expansion. 

Obviously, the zero-curvature form (\ref{zcurv}) is invariant under the gauge
transformation:
\be
\delta_n A_x(x,\lambda)=[\theta_n(x,\lambda),{\cal L}]
\label{gauge}
\ee
for $A_x$, and a similar one for $A_t$. A suitable choice for the gauge
parameter $\theta_n$ goes through the construction of the following object:
\be
Z^X(x,\lambda)=T(x,\lambda)XT(x,\lambda)^{-1} \virg  X=E,F,H
\label{zdress}
\ee
essentially the dressed generators of the underlying $A_1^{(1)}$ algebra. It
is obvious by construction that it satisfies:
\be
[\cL,Z^X(x,\la)]=0 ,
\label{resolv}
\ee
i.e. it is a resolvent of the Lax operator {\cal L}. As we shall see, this
property is important for the construction of a consistent gauge parameter.  In
fact, let us insert $T_{reg}$ as defined in (\ref{regsol}) in (\ref{zdress})
and then construct 
\be
\t^X_{-n}(x,\la)\equiv
(\la^{-n}Z^X(x,\la))_-=\sum^{n-1}_{k=0}\la^{k-n}Z^X_k(x) , 
\label{gaugepar}
\ee
where the subscript -- (+) means that we restrict the series only to negative
(non-negative) powers of $\la$. One can show that due to (\ref{resolv}) the
r.h.s. of (\ref{gauge}) is of degree zero in $\lambda$ and therefore
$\theta_{-n}^X$ so constructed is a good candidate for a consistent gauge
parameter. There is one more consistency condition we have to impose due to
the explicit form of $A_x$ (\ref{lax}), namely $\delta A_x$ should be
diagonal: 
\be \d^X_{-n}A_x=H\d^X_{-n}\phi' .
\label{selfcons}
\ee
This implies restrictions on the indices of the transformations: it happens
that one must take even ones for $X=H$ ( $\theta_{-2n}^H$ ) and odd ones for 
$X=E$ or $F$ ( $\theta_{-2n-1}^{E,F}$ ). The first transformations read
explicitly:
\ba
\delta^E_{-1} \phi'(x) &=&  e^{2\phi(x)}          \nonumber  \\
\delta^F_{-1} \phi'(x) &=&  - e^{-2\phi(x)}    \nonumber  \\
\delta^H_{-2} \phi'(x) &=& e^{2\phi(x)} \int_0^x dy e^{-2\phi(y)} +
e^{-2\phi(x)} \int_0^x dy 
e^{2\phi(y)} .
\label{che}
\ea
Note that they are essentially non-local (this is true also for the higher
ones). The algebra these transformations close is the  (twisted) Borel
subalgebra of $A_1^{(1)}$. Therefore the remaining ones can be found
from(\ref{che}) by commutation.

At this point we want to make an important observation. Consider the KdV
variable $x$ as a {\it space direction} $x_-$ of some more general system (and
$\p_-\equiv \p_x$ as a space derivative). Introduce the {\it time} variable
$x_+$and the corresponding evolution defining: 
\be
\p_+\equiv (\d^E_{-1}+\d^F_{-1}).
\label{timeevol}
\ee
It is then obvious from (\ref{che}) that the equation of motion for $\phi$
becomes:
\be
\p_+\p_-\phi=2\sinh(2\phi) \virg (or \spz 2sin(2\phi)\spz if\spz
\phi\rightarrow i\phi) \label{sgeq}
\ee
i.e. the sine-Gordon equation! We consider this observation very important
since it provides a {\it global} introduction of sine-Gordon dynamics in the
KdV  system - a fact which was not known before.

The construction of the gauge parameter in the asymptotic case goes along the
same line as above \cite{FS0}. Explicitly, the asymptotic expansion of the
transfer matrix is given by:
\be 
T(x,\la)_{asy}=KG(x,\la)e^{-\int_0^x dy D(y)},
\label{asyexp}
\ee
where
$K =\frac{\sqrt{2}}{2}\left(\begin{array}{cc}  1 & 1 \\
                                              1 & -1 \end{array}\right)$ and
\be
D(x,\la)=d(x,\la)H , \spz
d(x,\la)=\sum_{k=-1}^{\infty}\la^{-k}d_k(x)  
\label{lod}
\ee
($d_{2k+1}$ are the conserved densities). $G$ is given by:
\be
G(x,\la)=\buno + \sum_{j=1}^{\infty}\la^{-j}G_j(x) ,
\label{defg}
\ee
where $G_j(x)$ are off-diagonal matrices with entries $(G_j(x))_{12}=g_j(x)$
and $(G_j(x))_{21}=(-1)^{j+1}g_j(x)$ ( see \cite{FS0} ).

The resolvent (\ref{zdress}) where now $T=T_{asy}$ and $X=H$ is an infinite
series in negative powers of $\lambda$:
\be
Z(x,\la)=\sum_{k=0}^\infty \la^{-k}Z_k(x)
\label{zasy}
\ee
and obviously satisfies (\ref{resolv}) by construction. The coefficients in
(\ref{zasy}) are given by:
\be 
Z_{2k}(x)=b_{2k}(x)E+c_{2k}(x)F \virg Z_{2k+1}(x)=a_{2k+1}(x)H . 
\label{solasy} 
\ee 
A suitable gauge parameter in this case is constructed as:
\be
\theta_{n}(x,\la)=(\la^{n}Z(x;\la))_+=\sum_{j=0}^{n}
\la^{n-j} Z_j(x)  
\label{asypar}
\ee 
and the additional consistency condition (\ref{selfcons}) implies that now the
indices should be odd ( $\theta_{2n+1}$ ). It happens that these
transformations coincide exactly with the commuting higher mKdV flows (or mKdV
hierarchy):
\be 
\delta_{2k+1}\phi'(x)=\partial a_{2k+1}(x)
\label{mkdvh} 
\ee 
and are therefore local in contrast with the regular ones.
It turns out that the other entries of the resolvent $b_{2n}(x)$ are exactly
the conserved densities \cite{FS1}, namely: 
\be
\d_{2k+1}\phi'(x)=\{I_{2k+1},\phi'(x)\} \virg I_{2k-1}=\int_0^L dx b_{2k}(x).
\label{bdens}
\ee
They differ from $d_{2k+1}$ (\ref{lod}) by a
total derivative. For example:
\ba
b_2 &=& -d_1+\half \phi'' ,\nn \\
b_4 &=& {3\over 4}d_3+\p({7\over 32}\phi''\phi'+{1\over 16}(\phi')^3+{1\over
16}\phi''') \spz etc.
\label{bd}
\ea
Finally, let us note that it can be shown that these two kind of symmetries 
(regular and asymptotic) commute with each other. In this sence the non-local
regular transformations provide a true symmetry of the KdV hierarchy.

\subsection{Generalization - Virasoro Symmetry}

Now, let us explain how the Virasoro symmetry appears in the KdV
system \cite{FS}.
The main idea is that one can dress not only the generators of the underlying 
$A_1^{(1)}$ algebra but also an arbitrary differential operator in the spectral
parameter. We take for example $\lambda^{m+1}\partial_{\lambda}$ which, as it is well 
known, are the vector fields of the diffeomorphisms of the unit circumference
and close a Virasoro algebra. Then we proceed in the same way as above.

The analog of our basic object (\ref{zdress}) now is:
\be
Z^V(x,\la)=T(x,\la)\p_\la T(x,\la)^{-1} .
\label{zvir}
\ee
It is clear that $Z^V$ has again the property of being a resolvent for the Lax
operator,  i.e. it satisfies (\ref{resolv}), which was one of the requirements
for constructing a good gauge parameter. Let us consider first the regular
case, i.e. take  $T=T_{reg}$ in (\ref{zvir}):
\be
Z_{reg}^V(x,\la)=\sum_{n=0}^\infty \la^nZ_n(x)+\p_\la .
\label{zvirreg}
\ee
It is clear that $Z^V(x,\lambda)$ is a differential operator in $\lambda$ in
this case.  Following the same reasoning as before we construct the gauge
parameter as:
\be
\theta_{-m}^V(x,\la)=(\la^{-m}Z_{reg}^V(x,\la))_- \virg m>0 .
\label{tetavirreg}
\ee
Then the additional condition (\ref{selfcons}) imposes that the indices of the
transformation should be  even $m=2n$. The first nontrivial examples are given
by: 
\ba
\delta_{-2}^V \phi' &=&  e^{2\phi(x)} \int_0^x dy e^{-2\phi(y)} -
e^{-2\phi(x)} \int_0^x dy e^{2\phi(y)}=e^{2\phi(x)}B_1-e^{-2\phi(x)}C_1
\nonumber  \\ 
\delta_{-4}^V  \phi' &=&  e^{2\phi(x)}(3B_3(x)-A_2(x)B_1(x)) -
e^{-2\phi(x)}(3C_3(x)-D_2(x)C_1(x)) \nonumber  \\
\delta_{-6}^V  \phi' &=& e^{2\phi(x)} (5B_5(x)-3A_4(x)B_1(x)+A_2(x)B_3(x))
\nonumber  \\
&-& e^{-2\phi(x)}(5C_5(x)-3D_4(x)C_1(x)+D_2(x)C_3(x)) , 
\label{frd's} 
\ea
where $A_i$, $B_i$, $C_i$, $D_i$ were defined in (\ref{regmatrix}).
They have a form very similar to that of $\delta^H_{-2n}$ but nevertheless it
can be shown  \cite{FS1} that they indeed close a (half) Virasoro algebra.
Note that these transformations are essentially non-local, so we obtained a
very  nontrivial realisation of the Virasoro algebra in terms of vertex
operators.

Let us now consider the asymptotic case, i.e. take $T=T_{asy}$ in (\ref{zvir}):
\be
Z_{asy}^V(x,\la)=\sum_{n=0}^\infty \la^{-n}Z_n(x)+\p_\la .
\label{zvirasy}
\ee
The coefficients of the above expansion have the general form:
\be
Z_{2n}=\beta_{2n}E+\gamma_{2n}F ,\spz  Z_{2n+1}=\alpha_{2n+1}H \nonumber ,
\ee
where for example $\beta_0=x=\gamma_0$, $\alpha_1=2xg_1$,
$\beta_2=-xb_2-g_1+\int^x_0 d_1$, $\gamma_2=-xc_2+g_1+\int^x_0 d_1$ etc. .
Again, the suitable gauge parameter is defined by:
\be
\theta^V_{m}(x,\la)=(\la^mZ^V_{asy})_+=\sum_{n=0}^{m+1}\la^{m+1-n}Z_n +\dla
\virg m\ge 0 
\label{tetavirasy}
\ee
and the self-consistency condition implies that the indices should be even in
this case too.  Actually, the first transformation:
\be
\d_0^V\phi'(x)=(x\p+1)\phi'(x)
\label{deltazero}
\ee
is exactly the scale transformation - it counts the dimension (or level). 
The first non-trivial examples are:
\ba
\delta_{2}^V  \phi' &=& 2xa_3'-(\phi')^3+{3\over 4}\phi'''
+2a'_1\int^x_0 d_1, \nn \\ 
\delta_{4}^V  \phi' &=& 2xa_5'+(\phi')^5-{5\over 2}\phi'''(\phi')^2-{27\over
8} (\phi'')^2\phi'+{5\over 16}\phi^V +\nn \\
                    &+&2a_3'\int_0^x d_1 +6a_1'\int_0^x d_3 .
\label{deltatwo}
\ea
We note that these depend explicitly on $x$ and are quasi-local (they
contain some indefinite  integrals). For further reference we presented the
integrands in (\ref{deltatwo}) explicitly in terms of the entries of the basic
objects $T(x,\lambda)$ and $Z(x,\lambda)$, defined in (\ref{lod},\ref{solasy}).
 Furthermore, one
can find the transformation of the resolvent and therefore the transformation
of the conserved densities $\delta_{2k}b_{2n}(x)$. In particular the first
nontrivial transformations of the KdV variable $u=b_2$ read: \ba
\delta_2^Vb_2 &=& \delta_2^V u= 2xb_4'+u''-2u^2-{1\over 2}u'\int^x_0u , \nn \\
\delta_4^Vb_2 &=& \delta_4^V u= 2xb_6'+2u^3+3uu''+{17\over 8}(u')^2+{3\over
8}u^{IV} + \nn \\
              &+& u'\int_0^x b_4 +b_4'\int_0^x u .                             
\label{deltau} 
\ea 
It can be easily
shown that the asymptotic transformations also close (half) Virasoro algebra. 
A very non-trivial question concerns the commutation relations between the
``negative'' and ``positive'' parts so constructed in view of their completely
different nature. Nevertheless it can be shown \cite{FS1} that,  contrary to
what happened between the proper dressing transformations and the m-KdV
hierarchy, in this case they close a whole Virasoro algebra: \be
[\delta^V_{2m},\delta^V_{2n}]=(2m-2n)\delta^V_{2m+2n}  \virg  m,n \in {\Bbb Z}
. \label{vir} \ee We want to stress that the Virasoro symmetry just
cconstructed is a dynamical one and has  nothing to do with the space-time
Virasoro symmetry of CFT. Actually, it is well known that one can consider the
KdV system as a classical limit of the latter. So we expect that after
quantization this dynamical symmetry should be present in CFT. It is
interesting to investigate its significance, in particular if CFT could be
solved by using this symmetry alone.

\resection{Soliton Solutions of (m)KdV theory}

\subsection{(m)KdV Solitons}

We would like now to restrict the Virasoro symmetry to the soliton solutions
of the (m)KdV theory. One can expect that in this case it simplifies
considerably. There is also another reason for this restriction. It was shown
in \cite{BB} that one can reconstruct certain form-factors of sine-Gordon
theory by directly quantizing the soliton solutions. Moreover, it happens
that a kind of null-vectors appear in the theory \cite{BBS}, leading to
integral equations for the form-factors. It is intriguing to understand the
r\^ole that the Virasoro symmetry just described is playing in all these
constructions.

We start with a brief description of the well known soliton solutions of
(m)KdV. They are best expressed in terms of the so-called {\it
tau-function}. In the case of N-soliton solution of (m)KdV it has the form:
\be
\tau(X_1,...,X_N| B_1,...,B_N)=\det(1+V)
\label{tau}
\ee
where $V$ is a matrix:
\be
V_{ij}=2{B_iX_i(x)\over B_i+B_j} \virg i,j=1,...,N.
\label{vmatr}
\ee
The m-KdV field is then expressed as:
\be
e^\phi={\tau_-\over \tau_+} ,
\label{phiintau}
\ee
where:
\be
\tau_\pm(x)=\tau(\pm X(x)|B)
\label{taupm}
\ee
and $X_i(x)$ is simply given by:
\be
X_i(x)=X_i\exp(2B_ix).
\ee
The variables $B_i$ and $X_i$ are the parameters describing the solitons:
$\beta_i=\log B_i$ are the so-called rapidities and $X_i$ are related to the
positions. The integrals of motion, restricted to the N-soliton solutions have
the form 
\be
I_{2n+1}=\sum_{i=1}^N B_i^{2n+1} \virg n\ge0 .
\label{inmotb}
\ee
It is well known that (m)KdV admits a non-degenerate symplectic structure. One
can find the corresponding Poisson brackets between the basic variables $B_i$
and $X_i$ \cite{BB0}. The (m)KdV flows are then generated by (\ref{inmotb})
via
\be
\delta_{2n+1}*=\{\sum_{i=1}^N B^{2n+1},*\} \virg n\ge0 .
\label{solkdv}
\ee

\subsection{Analytical Variables}

Our final goal is the quantization of solitons and of the Virasoro symmetry. It
was argued in \cite{BB} that this is best performed in another set of
variables $\{A_i,B_i\}$. The latter are the soliton limit of certain variables
describing the more general quasi-periodic finite-zone solutions of (m)KdV. In
that context $B_i$ are the branch points (i.e. define the complex structure) of
the hyperelliptic Riemann surface describing the solution and $A_i$ are the
zeroes of the so-called Baker-Akhiezer function defined on it. In view of the
nice geometrical meaning of these variables they were called analytical
variables in \cite{BB}.

Explicitly, the change of variables is given by:
\be
X_j\prod_{k\ne j}{B_j-B_k\over B_j+B_k}=\prod_{k=1}^N{B_j-A_k\over B_j+A_k} 
\virg j=1,...,N .
\label{analytic}
\ee
 The non-vanishing Poisson
brackets expressed in terms of these new variables 
take the form: 
\be
\{A_i,B_j\}={\prod_{k\ne i}(B_j^2-A_k^2)\prod_{k\ne j}(A_i^2-B_k^2)\over
\prod_{k\ne i}(A_i^2-A_k^2)\prod_{k\ne j}(B_j^2-B_k^2)}(A_i^2-B_j^2).
\label{poisson}
\ee
The corresponding tau-functions have also a very compact form in terms of the
analytical variables:
\ba
\tau_+ &=& 2^N\prod_{j=1}^N B_j\{{\prod_{i<j}(A_i+A_j)\prod_{i<j}(B_i+B_j)\over
\prod_{i,j}(B_i+A_j)}\} \nonumber \\
\tau_- &=& 2^N\prod_{j=1}^N A_j\{{\prod_{i<j}(A_i+A_j)\prod_{i<j}(B_i+B_j)\over
\prod_{i,j}(B_i+A_j)}\}.
\label{anlitictau}
\ea
Therefore, from the explicit form of the m-KdV field in terms of the
tau-functions (\ref{phiintau}) we obtain the following very simple expression:
\be
e^\phi\equiv{\tau_-\over\tau_+}=\prod_{j=1}^N{A_j\over B_j}.
\label{analiticphi}
\ee
The equation of motion of the $A_i$ variable is given by:
\be
\p_xA_i\equiv \d_1A_i=\{I_1,A_i\}=\prod_{j=1}^N(A_i^2-B_j^2)\prod_{j\ne
i}{1\over(A_i^2-A_j^2)}.
\label{eqmot}
\ee
One can verify that, as a consequence, the usual KdV variable $u$ is expressed
as: \be
b_2\equiv u=\half (\phi')^2+\half \phi''=\sum_{j=1}^NA_j^2-\sum_{j=1}^NB_j^2.
\label{analitcu}
\ee
One can restrict also the higher KdV flows to the soliton solutions. For
example it is clear from (\ref{solkdv}) that
\be
\d_{2n+1}B_i=0 \virg n\ge 0 .
\ee
This is a reminiscence of the fact that the KdV flows do not change the complex
structure of the hyperelliptic surface describing the finite-zone solution
\cite{GO}. The variation of the $A_i$ variables can be easily computed as :
\be
\d_{2n+1}A_i=\{I_{2n+1},A_i\} \virg n\ge0
\ee
using the Poisson brackets (\ref{poisson}).

\subsection{Virasoro Symmetry of the Soliton Solutions}

Now, we want to restrict the Virasoro symmetry of (m)KdV constructed above to
the case of soliton solutions. In this section we shall be only interested in
the positive part of the latter. The transformation of the rapidities can be
easily deduced as a soliton limit of the Virasoro action on the finite-zone
solutions described in \cite{GO}: \be
\d_{2n}B_i=B_i^{2n+1} \virg n\ge 0 ,
\label{btransf}
\ee
i.e. the Virasoro action changes the complex structure (because of that it's
often called non-isospectral symmetry).
What remains is to obtain the transformations of the $A_i$ variables. We found
it quite difficult to deduce them as a soliton limit of the corresponding
transformations of  \cite{GO}. Instead, we propose here another approach.
Namely, we use the transformation of the fields $\d_{2n}\phi$, $\d_{2n}\phi'$,
$\d_{2n}u$ {\it etc.} which we found before, restricted to the soliton
solutions using (\ref{analiticphi},\ref{analitcu}). The problem is
simplified by the fact that the Virasoro algebra is freely generated ,
i.e. we need to compute only the $\d_0$, $\d_2$ and $\d_4$ transformations, the
remaining ones are then obtained by commutation. In practice, we perform the
computation for the first few cases of $N=1,2,3$ solitons and then proceed by
induction.

Let us make an important observation. As we have stressed, the transformation
of the basic objects in the field theory of (m)KdV are quasi-local -- they
contain certain indefinite integrals. It happens that the corresponding
integrands become total derivatives when restricted to the soliton solutions.
For example: \ba
b_2 &\equiv &u=\p_x\sum_{i=1}^NA_i(x) , \nonumber \\
b_4 &=& \p_x\sum_{i=1}^N A_i^3-\half u'\equiv \p_x[\sum_{i=1}^N(A_i^3-\half
\p_xA_i)].
\label{totderiv}
\ea
Therefore the Virasoro transformations become {\it local} when restricted to
the soliton solutions! The calculation is straightforward but quite tedious
so we present here only the final result:
\ba
\d_0 A_i &=& (x\p_x+1)A_i , \nonumber \\
\d_2 A_i &=&{1\over 3}x\d_3A_i+A_i^3-(\sum_{j=1}^N A_j)\p_xA_i , \nonumber \\
\d_4 A_i &=&{1\over 5}x\d_5A_i+A_i^5-\{\sum_{j\ne
i}A_i(A_i^2-A_j^2)+\sum_{j=1}^NA_j \sum_{k=1}^N B_k^2\}\p_xA_i ,
\label{result}
\ea
where the KdV flows read explicitly:
\ba
{1\over 3}\d_3A_i &=& (\sum_{j=1}^N B_j^2-\sum_{k\ne i}A_k^2)\p_xA_i ,
\nonumber \\ 
{1\over 5}\d_5A_i &=& (\sum_{j=1}^N B_j^4-\sum_{k\ne i}A_k^4)\p_xA_i -
\sum_{j\ne i}(A_i^2-A_j^2)\p_xA_i\p_xA_j.
\label{kdvonsol}
\ea
As we already mentioned, the remaining transformations can be obtained by
commutation, for example:
\be
2\d_6A_i=[\d_4,\d_2]A_i \virg etc.
\ee

\resection{Virasoro Symmetry of Sine-Gordon Solitons}

\subsection{From (m)KdV to Sine-Gordon Solitons}

Now we pass to the most important part of our paper. We would like to extend
the construction presented above in (m)KdV theory to the case of sine-Gordon.
For this purpose one has to find a way of extending the mKdV dynamics up to the
sine-Gordon one. It is to some extent known how this can be done in the case
of the soliton solutions \cite{BB}. The idea is close to what we proposed
before directly in the field theory of (m)KdV. Namely, let us consider the KdV
variable $x$ as a {\it space} variable of some more general system and call it
$x_-$ ( and $\p_-\equiv \p_x$ correspondingly ). We would like to introduce a
new {\it time} variable $x_+$ and the corresponding time dynamics. In the case
of the N - soliton solutions the latter is generated by the Hamiltonian:
\be
I_{-1}=\sum_{i=1}^NB_i^{-1}
\ee
( essentially the inverse power of the momentum ) so that the {\it time} flow
is given by:
\be
\p_+*=\d_{-1}*=\{I_{-1},*\}
\label{timeflow}
\ee
using again the Poisson brackets (\ref{poisson}). In particular:
\be
\p_+A_i=\prod_{j=1}^N{A_i^2-B_j^2)\over B_j^2}\prod_{j\ne i}{A_j^2\over
(A_i^2-A_j^2)}.
\label{plusa}
\ee
One can check, using (\ref{analiticphi}, \ref{eqmot}), that with this
definition the resulting equation for the field $\phi$ is
\be
\p_+\p_-\phi=2\sinh (2\phi)
\label{sinhg}
\ee
or under the change $\phi\rightarrow i\phi$:
\be
\p_+\p_-\phi=2\sin (2\phi)
\label{sing}
\ee
i.e. the sine-Gordon equation. We were not able to establish at the moment a
direct relation between the way of extending the mKdV dynamics to the
sine-Gordon one directly in the field theory (\ref{timeevol}) and in the case
of N-soliton solution just described (\ref{timeflow}). We hope to answer this
important question elsewhere \cite{35}. In a similar manner one can introduce
the rest of the sine-Gordon Hamiltonians:
\be
I_{-2n-1}=\sum_{i=1}^NB^{-2n-1}_i \virg n\ge 0 .
\label{negcharges}
\ee
They generate the ``negative KdV flows'' via the Poisson brackets
(\ref{poisson}): 
\ba
\d_{-2n-1}B_i &=& 0 , \nonumber \\
\d_{-2n-1}A_i &=& \{I_{-2n-1},A_i\} \virg n\ge 0 .
\label{negkdv}
\ea

\subsection{Negative Virasoro flows}

Now we arrive at the {\bf main conjecture} of this paper. Having in mind the
symmetric role the derivatives $\p_-$ and $\p_+$ are playing in the sine-Gordon
equation {\it we would like to suppose that one can obtain another half
Virasoro algebra by using the same construction as above but with $\p_-$
interchanged with $\p_+$}!

So let us define as before:
\be
\d_{-2n}B_i=-B_i^{-2n+1} \virg n\ge 0
\label{negb}
\ee
(note the additional -- sign in the r.h.s. which is needed for
the self-consistency of the construction). Following our conjecture we
construct the negative flows of the $A_i$ variable in the same way as before
but with the change $\p_-\rightarrow\p_+$. We have for example:
\ba
\d_{-2}\phi &=& x_+(2a_3^+)+b_2^+-2a_1^+\int^{x_+}_0b_2^+ , \nn \\
\d_{-2}b_2^- &\equiv& \d_{-2}u ={1\over 3}x_+\d_{-3}u
+(\p_+\phi-\int^{x_+}_0b_2^+)\p_-e^{2\phi} , 
\label{negu}
\ea
where $\d_{-3}u\equiv \{I_{-3},u\}$ {\it etc.} In (\ref{negu}) the + subscript
means that we take the same objects as defined in (\ref{lod}, \ref{defg},
\ref{solasy}) but with $\p_-$ changed by $\p_+$. For example:
\ba
b_2^+ &=& \half (\p_+\phi)^2 +\half \p_+^2\phi ,\nonumber \\
a_3^+ &=& -{1\over 4}(\p_+\phi)^3 +{1\over 8} \p_+^3\phi \spz etc.
\label{pluses}
\ea
At this point we
want to make an important remark. Very non-trivially, it happens again that the
integrands in the expressions (\ref{negu}) and similar become total derivatives
when restricted to the N-soliton solutions. So that again the (negative)
Virasoro symmetry is {\it local} in the case of solitons! We present below the
first examples of this phenomenon:
\ba
b_2^+ &=& \p_+\{\sum_{i,j=1}^N{A_iA_j\over
B_i^2B_j^2}\sum_{i=1}^NA_i-\p_-\sum_{i,j=1}^N{A_iA_j\over B_i^2B_j^2}\},
\nonumber \\ 
b_4^+ &=& \p_+\{\sum_{i,j=1}^N{A_iA_j\over
B_i^4B_j^4}\sum_{i=1}^NA_i^3+ b_2^-\p_-\sum_{i,j=1}^N{A_iA_j\over
B_i^4B_j^4}- \nonumber \\       
&-& \p_-b_2^-\sum_{i,j=1}^N{A_iA_j\over
B_i^4B_j^4}\} \spz etc. 
\label{totderplus}
\ea
We then proceed as in the case of the positive Virasoro flows, i.e. we restrict
the transformations of the fields thus obtained to the soliton solutions.
As we explained, it is enough to find only the first transformations
$\d_{-2}A_i$ and $\d_{-4}A_i$ and the remaining ones are found by
commutation. Following our approach we do the computation explicitly in the
case of $N=1,2,3$ solitons and then proceed by induction. The exact
calculation will be presented in a forthcoming paper \cite{35}, here we give
the final results only:
\ba
\d_{-2}A_i &=& \frac{1}{3} x_+\d_{-3}A_i
-A_i^{-1}-(\sum_{j=1}^NA_j^{-1})\p_+A_i, \nn \\ 
\d_{-4}A_i &=& \frac{1}{5} x_+\d_{-5}A_i
-A_i^{-3}- \nn \\            
&-&\{\sum_{j\neq i}^N{1\over A_i}({1\over
 A_i^2}-{1\over A_j^2})+\sum_{j=1}^N{1\over A_j}\sum_{k=1}^N{1\over
 B_k^2}\}\p_+A_i ,
\ea
where as before $\d_{-3}A_i=\{\sum_{j=1}^NB_j^{-3},A_i\}$ etc. As stated
above, we then can compute $2\d_{-6}A_i=[\d_{-2},\d_{-4}]A_i$ etc.

\subsection{The Algebra}

Now, we come to the important problem of the commutation relations between the
two half Virasoro algebras so constructed. This is a non-trivial question in
view of the different way we obtained them. In fact, it is clear that, by
construction, the positive (negative) Virasoro flows commute with the
corresponding $\p_-$ ( $\p_+$ ) derivatives:
\ba
[\d_{2n},\partial_-]A_i &=& 0 , \nn \\
\vspace{0.5cm}
[\d_{-2n},\partial_+]A_i &=& 0 \virg n\ge 0 .
\label{commut}
\ea
It is easy to see that this is not true for the ``cross commutators''.
Actually, one finds in this case:
\ba
[\d_{2n},\p_+]A_i &=& -\d_{2n-1}A_i , \nn \\
\vspace{0.5cm}
[\d_{-2n},\p_-]A_i &=& -\d_{-2n+1}A_i \virg n\ge 0 .
\label{noncommut}
\ea
It is clear that we are interested in a {\it true symmetry} of the sine-Gordon
theory. We must therefore obtain transformations that commute with the $\p_-$
and $\p_+$ flows and as a consequence with the corresponding Hamiltonians. It
is obvious from (\ref{commut},\ref{noncommut}) that this is acheeved by a
simple modification of the flows, i.e. let us define:
\ba
\d_{2n}' &=& \d_{2n}-x_+\d_{2n-1} , \nn \\
\d_{-2n}' &=& \d_{-2n}-x_-\d_{-2n+1} \virg n\ge 0 .
\label{modtransf}
\ea
Then, for the modified transformation we obtain:
\be
[\d_{2n}',\p_{\pm}]A_i=0 \virg n\in {\Bbb Z}.
\ee
Finally, one can show that, with this modification, the commutation relations
between the positive and negative parts of the transformations close exactly
the whole Virasoro algebra:
\be
[\d_{2n}',\d_{2m}']A_i=(2n-2m)\d_{2n+2m}' A_i \virg n,m\in {\Bbb Z}.
\label{primevir}
\ee 

\resection{Conclusions and discussion}

To summarise, we presented in this paper a construction of a Virasoro symmetry
of the sine-Gordon theory. This is acheeved by extending the corresponding
symmetry of the (m)KdV theory \cite{FS}. Actually, we found it easier to work
rather with the N-soliton solutions of the latter. Then, following \cite{BB},
we introduced a {\it time} coordinate $x_+$ and the corresponding flow $\p_+$
in addtion to the {\it space} variable $x_-$ of KdV which leads to the
sine-Gordon equation. The {\bf main idea} is to make the same kind of
construction as for KdV but interchanging the $\p_-$ with the $\p_+$
derivative. This change results in what we called ``negative'' Virasoro flows
which complete the ``positive'' ones coming from the original KdV to the whole
Virasoro algebra. We showed also that after a certain modification, needed to
obtain a true symmetry of the theory, they close the whole Virasoro algebra.

Actually, in this paper we obtained the infinitesimal transformations of the
variables describing the N-soliton solutions. It is intriguing to find the
corresponding conserved charges $J_{2n}(A_i, B_i)$. This is important in view
of the quantization of the classical constructions presented here. Besides, 
in their commutation
relations one
can obtain a possible existence of a central extension which cannot be discovered 
 in the case of the infinitesimal
transformations. 

As already mentioned, we are interested of course in the quantum sine-Gordon
theory. In the case of solitons there is a standard procedure, a kind of 
canonical quantization of the N-soliton solutions. In fact, let us introduce,
following \cite{BB}, the canonically conjugated variables to the {\it
analytical variables} $A_i$:
\be
P_j=\prod_{k=1}^N{B_k-A_j\over B_k +A_j} \virg j=1...,N
\ee
(in the variables $\{P_j,A_j\}$ the corresponding simplectic structure is
diagonal). In these variables one can perform a kind of canonical quantization
of the N-soliton system introducing the deformed commutation relations between
the operators $A_i$ and $P_i$ :
\ba
P_jA_j &=& q A_jP_j , \nn \\
P_kA_j &=& A_jP_k \spz for \spz k\ne j ,
\ea
where $q$ is related to the sine-Gordon coupling constant: $\exp (i\xi)$,
$\xi={\pi\gamma\over \pi-\gamma}$. It is very intriguing to understand how the
Virasoro symmetry is deformed after the quantization! 

Another important problem is the construction of the Virasoro symmetry
directly in the sine-Gordon field theory. We explained above how this can be
done using the proper dressing transformations. It happens that the positive
Virasoro flows commute with the {\it time} sine-Gordon flow $x_+$ introduced in
this way. We expect that the negative Virasoro symmetry can be constructed
following the same approach we presented in this paper for the solitons.

Finally, it will be very interesting to understand the r\^ole this Virasoro
symmetry is playing in the sine-Gordon theory. As we mentioned, it was shown in
\cite{BB} that certain form-factors can be reconstructed by a suitable
quantization of the N-soliton solutions. One can expect that the Virasoro
symmetry imposes some constraints leading to certain equations for the
form-factors (or correlation functions in the case of field theory). We would
like to note in this respect the article \cite{BBS} where a kind of
null-vector constraints were derived in the sine-Gordon theory. The
corresponding construction is closely related to the finite-zone solutions. As
we mentioned the Virasoro symmetry we presented here has a natural action on
such solutions. It is natural to expect that some relation exists between this
symmetry and the null-vectors of \cite{BBS}. We will return to all these
problems in a forthcoming paper \cite{35}.

\vspace{1cm}

{\bf Acknowledgments} - We are indebted to E. Corrigan, V. Fateev, G. Mussardo,
G. Sotkov and Al. Zamolodchikov for discussions and interest in this work.
D.F. thanks the I.N.F.N.--S.I.S.S.A. for financial support. M.S. acknowledges
S.I.S.S.A.  and the University of Montpellier -2 for the warm hospitality over
part of this work. This work has been realized through partial financial
support of TMR Contract ERBFMRXCT960012.

\end{document}